\documentclass[aps,prl,twocolumn,showpacs,superscriptaddress]{revtex4-1}

\usepackage[final]{graphicx}
\usepackage{amsmath}
\usepackage{verbatim}
\usepackage{color}
\usepackage{ulem}
\usepackage[colorlinks,
linkcolor=blue,
citecolor=blue,
urlcolor=blue]{hyperref}

\begin{document}

\preprint{}

\title{Anti-Stokes scattering and Stokes scattering of stimulated Brillouin scattering cascade in high-intensity laser-plasmas interaction}

\author{Q. S. Feng} 
\affiliation{HEDPS, Center for
	Applied Physics and Technology, Peking University, Beijing 100871, China}
\author{Z. J. Liu} 
\affiliation{HEDPS, Center for
	Applied Physics and Technology, Peking University, Beijing 100871, China}
\affiliation{Institute of Applied Physics and Computational
	Mathematics, Beijing, 100094, China}

\author{C. Y. Zheng} \email{zheng\_chunyang@iapcm.ac.cn}

\affiliation{HEDPS, Center for
	Applied Physics and Technology, Peking University, Beijing 100871, China}
\affiliation{Institute of Applied Physics and Computational
	Mathematics, Beijing, 100094, China}
\affiliation{Collaborative Innovation Center of IFSA (CICIFSA) , Shanghai Jiao Tong University, Shanghai, 200240, China}

\author{C. Z. Xiao}
\affiliation{School of Physics and Electronics, Hunan University, Changsha 410082, China}

\author{Q. Wang}
\affiliation{HEDPS, Center for
	Applied Physics and Technology, Peking University, Beijing 100871, China}

\author{L. H. Cao} 
\affiliation{HEDPS, Center for
	Applied Physics and Technology, Peking University, Beijing 100871, China}
\affiliation{Institute of Applied Physics and Computational
	Mathematics, Beijing, 100094, China}
\affiliation{Collaborative Innovation Center of IFSA (CICIFSA) , Shanghai Jiao Tong University, Shanghai, 200240, China}

\author{X. T. He} \email{xthe@iapcm.ac.cn}
\affiliation{HEDPS, Center for
	Applied Physics and Technology, Peking University, Beijing 100871, China}
\affiliation{Institute of Applied Physics and Computational
	Mathematics, Beijing, 100094, China}
\affiliation{Collaborative Innovation Center of IFSA (CICIFSA) , Shanghai Jiao Tong University, Shanghai, 200240, China}


\date{\today}

\begin{abstract}
The anti-Stokes scattering and Stokes scattering in stimulated Brillouin scattering (SBS) cascade have been researched by the Vlasov-Maxwell simulation. In the high-intensity laser-plasmas interaction, the stimulated anti-Stokes Brillouin scattering (SABS) will occur after the second stage SBS rescattering. The mechanism of SABS has been put forward to explain this phenomenon. And the SABS will compete with the SBS rescattering to determine the total SBS reflectivity. Thus, the SBS rescattering including the SABS is an important saturation mechanism of SBS, and should be taken into account in the high-intensity laser-plasmas interaction.
	
\end{abstract}


\maketitle

Backward stimulated Brillouin scattering (SBS), i.e., the Stokes scattering, is a three-wave interaction process where an incident electromagnetic wave (EMW) decays into a backscattered EMW and a forward propagating ion-acoustic wave (IAW). Backward SBS leads to a great energy loss of the incident laser and is detrimental in inertial confinement fusion (ICF) \cite{He_2016POP,Glenzer_2010Science,Glenzer_2007Nature}. Therefore, SBS plays an important role in the successful ignition goal of ICF. Many mechanisms for the saturation of SBS have been put forward, including the creation of cavitons in plasmas \cite{Liu_2009POP_1, Weber_2005PRL, Weber_2005POP,Weber_2005POP_1}, frequency detuning due to particles trapping \cite{Froula_2002PRL,Giacone_1998POP,Vu_2001PRL,Albright_2016POP}, coupling with higher harmonics \cite{Bruce_1997POP, Rozmus_1992POP}, increasing linear Landau damping by kinetic ion heating \cite{Rambo_1997PRL, Pawley_1982PRL}, and so on.
However, if the pump light intensity is large enough, or the IAW Landau damping is low enough, it is possible for the scattered light to be scattered again. The multi-stage rescattering of SBS is called SBS cascade \cite{Speziale_1980POF,Liu_2012CPB,Robert_1981POF}. In this paper, the rescattering of SBS is observed and  found to be an important saturation mechanism of SBS in high-intensity laser region. Although the theoretical work on SBS rescattering \cite{Speziale_1980POF} gives a prediction of reduced SBS reflectivity under the assumption that the incident light is allowed to scatter only twice. In fact, the multiple SBS rescattering will occur in high-intensity laser region \cite{Liu_2012CPB, Robert_1981POF}, and different stage SBS rescatterings will have the different effects on the total reflectivity or the total transmitivity of SBS. This paper will give a detail analysis of each stage SBS rescattering evolution with time, and demonstrate the SBS rescatterings compete with each other to affect the reflectivity or the transmitivity.

In addition to the Stokes scattering in the SBS cascade, which is a common scattering mechanism of SBS, there exists a novel scattered light of a higher frequency than the pump light frequency.  This novel scattering is called stimulated anti-Stokes Brillouin scattering (SABS). The SABS is that a pump EMW couples with an IAW to produce a backward scattered EMW, the three waves satisfy the frequency and wave vector match conditions. The anti-Stokes Raman scattering was researched in the intense light interaction with gas \cite{Regnier_1973APL, Hickman_1988PRA}, liquid \cite{Manz_2004OC} and solid \cite{Kneipp_2000PRL}.   And the anti-Stokes scattering of SBS was researched in the intense light interaction with liquid \cite{Goldblatt_1968PRL} and solid \cite{Shin_2013NC}.  However, the research of anti-Stokes scattering in SBS cascade in the high-intensity laser-plasmas interaction has seldomly been reported. This paper gives an insight into the SABS in the high-intensity laser-plasmas interaction. We found that only when the second stage SBS rescattering (denoted SBS2) occurs, can the IAW generated by SBS2 couple with the pump light to generate a higher-frequency scattered light, which is the stimulated anti-Stokes Brillouin scattering (SABS) process. And the SABS can be even stronger than the third stage SBS rescattering (SBS3), thus SABS should be considered as a competition with other SBS rescattering. The total SBS reflectivity is the result of effects of SBS rescattering and SABS.

An one dimension in space and one dimension in velocity (1D1V) Vlasov-Maxwell code \cite{Liu_2009POP,Liu_2011POP,Liu_2012PPCF} is used to simulate the SBS cascade and SABS process. The electrons temperature is $T_e=5keV$. And the electrons density is $n_e=0.3n_c$, where $n_c$ is the critical density for the incident laser. In this paper, the density is larger than the quarter critical density, thus the stimulated Raman scattering does not occur. As a result, we can research the novel scattering in SBS cascade separately. The C plasma is taken as a typical example for it is common in ICF \cite{He_2016POP}. The Landau damping of the C plasmas is very low, thus the SBS cascade in C plasmas is easier to occur. The ions temperature is $T_i=T_e$. The linearly polarized laser wavelength $\lambda_0=0.351\mu m$ and the intensity varies. We can see that only when the pump light intensity reaches a certain value, such as $I_0=1\times10^{16}W/cm^2$, can the SBS cascade and SABS occur. Thus, the case in the intensity $I_0=1\times10^{16}W/cm^2$ will be taken as a typical example. To make the SBS increase more quickly, the seed light at the right boundary with a low intensity of $I_s=1\times10^{10}W/cm^2$ and a matching frequency $\omega_s=0.997\omega_0$. And in all of the cases in our simulations, the seed light is the same. The spatial scale is [0, $L_x$] discretized with $N_x=5000$ spatial grid points and spatial step $dx=0.2c/\omega_0$. And the spatial length is $L_x=1000c/\omega_0\simeq160\lambda_0$ with $2\times5\%L_x$ vacuum layers and $2\times5\%L_x$ collision layers in the two sides of plasmas boundaries. The incident laser propagates along the $x$ axis from the left to the right with outgoing boundary conditions. The strong collision damping layers are added into the two sides of the plasmas boundaries to damp the electrostatic waves such as IAWs at the boundaries. The velocity scale is discretized with $N_v=512$ grid points. The total simulation time is $t_{end}=1\times10^5\omega_0^{-1}$ discretized with $N_t=5\times10^5$ and time step $dt=0.2\omega_0^{-1}$.
 
 \begin{figure}[!tp]
 	\includegraphics[width=1\columnwidth]{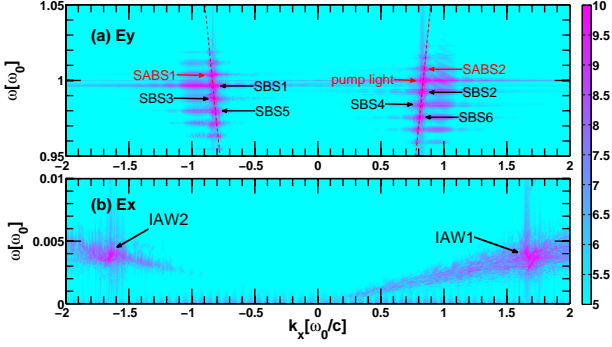}
 	
 	\caption{\label{Fig:w_k}(Color online) The dispersion relation of (a) transverse electric field $E_y$ and (b) longitudinal electric field $E_x$. The parameters are $n_e=0.3n_c, T_e=5keV, I_0=1\times10^{16}W/cm^2$ in C plasmas.}
 \end{figure}

Figure \ref{Fig:w_k} shows the dispersion relation of SBS cascade and stimulated anti-Stokes Brillouin scattering (SABS). The Stokes scattering in the SBS is commonly referred to as the stimulated Brillouin scattering, and the anti-Stokes scattering in the SBS is called SABS in this paper. In the first stage stimulated Brillouin scattering (SBS1), the pump light (EMW0) will resonantly decay into an IAW (denoted as IAW1 in Fig. \ref{Fig:w_k}(b)) and an inverse Stokes-scattered EMW (denoted as SBS1 in Fig. \ref{Fig:w_k}(a)), i.e., SBS1: $EMW0\rightarrow EMW1+IAW1$. If the scattered light EMW1 is strong enough and the Landau damping of the IAW is low enough, the second stage stimulated Brillouin scattering (SBS2) will occur, i.e., SBS2: $EMW1\rightarrow EMW2+IAW2$. Similarily, the third stage stimulated Brillouin scattering (SBS3) is: $EMW2\rightarrow EMW3+IAW3$. Thus, the stage $n$ of stimulated Brillouin scattering (SBSn) is: $EMW{(n-1)}[\vec{k}_{n-1}, \omega_{n-1}]\rightarrow EMWn[\vec{k}_{n}, \omega_{n}]+IAWn[\vec{k}_{IAWn}, \omega_{IAWn}]$, where $[\vec{k}_i, \omega_i]$ are the wave vector and the frequency of the corresponding waves. The matching condition of the waves in the stage $n$ ($n\geq 1$) of SBS is:
\begin{equation}
\label{Eq:waveVector}
\vec{k}_{n-1}=\vec{k}_n+\vec{k}_{IAWn},\quad \omega_{n-1}=\omega_n+\omega_{IAWn}.
\end{equation}
Because $\omega_{IAWn}\ll\omega_n$, and the direction of the wave vectors of the pump light and the scattered light are inverse for backward SBS discussed here, Eq. (\ref{Eq:waveVector}) can be written as:
\begin{equation}
\label{Eq: k_n}
k_n\simeq-k_{n-1}, \quad k_{IAWn}\simeq 2k_{n-1}.
\end{equation}
In addition to the Stokes scattering in the SBS cascade, there exist some novel scattered lights with higher frequencies than the pump light frequency as shown in Fig. \ref{Fig:w_k}(a). These novel scattered lights come from the SABS. The first stage anti-Stokes scattering process in SBS is that the pump light (EMW0) couples with the inverse strong IAW fluctuations (IAW2) to produce the higher-frequecy scattered light ($EMW_{-1}$ ), i.e., SABS1:
\begin{equation}
\begin{aligned}
&EMW0+IAW2\rightarrow EMW_{-1}.\\
&[\vec{k}_{0}, \omega_{0}]+[\vec{k}_{IAW2}, \omega_{IAW2}]=[\vec{k}_{-1}, \omega_{-1}].
\end{aligned}
\end{equation} 
 Therefore, the IAW2 must be produced before the SABS1 and the intensity of the pump light and IAW2 should be strong enough. If the anti-Stokes-scattered light ($EMW_{-1}$) produced in SABS1 is also strong enough, the second stage SABS will occur, i.e., SABS2:
\begin{equation}
\begin{aligned}
&EMW_{-1}+IAW1\rightarrow EMW_{-2}.\\
&[\vec{k}_{-1}, \omega_{-1}]+[\vec{k}_{IAW1}, \omega_{IAW1}]=[\vec{k}_{-2}, \omega_{-2}].
\end{aligned}
\end{equation} 

The dispersion relation of EMWn is:
\begin{equation}
\label{Eq: wn_kn}
\omega_n^2=\omega_{pe}^2+k_n^2c^2,
\end{equation}
where $c$ is the light speed in vacuum, $\omega_{pe}$ is the electrons plasmas frequency, $n=0$ represents the pump incident light, $n=1, 2, 3, ...$ represent the Stokes scattered lights, and $n=-1, -2$ represent the anti-Stoked scattered lights. In the single-species plasmas, the dispersion relation of the IAW is:
\begin{equation}
\label{Eq: w_IAWn}
\omega_{IAWn}=k_{IAWn}*c_s,
\end{equation}
the linear frequency, thus the sound velocity $c_s$, and Landau damping of IAW can be obtained by solving the equation\cite{Feng_2016POP,Feng_2016PRE,Chapman_2013PRL}:
\begin{equation}
\begin{aligned}
\label{Eq:Dispersion_C}
&\epsilon_L(\omega_{s},k_{IAW}=
2k_{0})=\\
&1+\sum_j^{species} \frac{1}{(k_{s}\lambda_{Dj})^2}(1+\xi_jZ(\xi_j))=0,
\end{aligned}
\end{equation}
where $j$ represents electrons or C ions,
$Z(\xi_j)=1/\sqrt{\pi}\int_{-\infty}^{+\infty}e^{-v^2}/(v-\xi_j)dv$ is the dispersion function,
and $\xi_j=\omega_{s}/(\sqrt{2}k_{IAW}*v_{tj})$ is complex and $\omega_{s}=Re(\omega_{s})+i\gamma_s$, $\gamma_s$ is the linear Landau damping of IAW; $v_{tj}=\sqrt{T_j/m_j}$, $\lambda_{Dj}=\sqrt{T_j/4\pi n_jZ_j^2e^2}$ is the thermal velocity and the Debye length of specie $j$. And $m_j, T_j, n_j, Z_j$ are the mass, temperature, density and charge number of specie $j$, respectively. 

From Eq. (\ref{Eq: wn_kn}), one can obtain the wave number of the pump incident light $k_0=0.8367\omega_0/c=0.1510\lambda_{De}^{-1}$ in the condition of $n_e=0.3n_c, T_e=5keV$. From Eq. (\ref{Eq: k_n}), the wave numbers of IAWs are: $k_{IAW1}=k_{IAW3}=...=k_{IAW(2n-1)}=2k_0$, and $k_{IAW2}=k_{IAW4}=...=k_{IAW(2n)}=-2k_0$. By solving Eq. (\ref{Eq:Dispersion_C}), the frequency of the IAW1 generated by the first stage SBS in C plasmas is $\omega_{IAW1}=Re(\omega_{s})=6.27\times10^{-3}\omega_{pe}=3.4\times10^{-3}\omega_0$, thus $c_s=\omega_{IAW1}/k_{IAW1}=0.0208v_{te}$. And the frequencies of IAWs are: $\omega_{IAW1}=\omega_{IAW3}=...=\omega_{IAW(2n-1)}=\omega_{IAW2}=\omega_{IAW4}=...=\omega_{IAW(2n)}$. Thus, we can think that $IAW1=IAW3=...=IAW(2n-1)$ and $IAW2=IAW4=...=IAW(2n)$. From Eq. (\ref{Eq:waveVector}), one can further obtain the frequencies of the scattered lights: $\omega_{n}=\omega_{n-1}-\omega_{IAWn}, (n=1, 2, 3, ...)$, i.e., $\omega_1=\omega_0-\omega_{IAW1}=0.9966\omega_0$, $\omega_{2}=\omega_{1}-\omega_{IAW2}=0.9932\omega_0$, $\omega_3=\omega_2-\omega_{IAW3}=0.9898\omega_0$, $\omega_4=\omega_3-\omega_{IAW4}=0.9864\omega_0$, ..., $\omega_n\simeq\omega_{0}-n*\omega_{IAW1}=(1-3.4\times10^{-3}*n)\omega_0$. Through simultaneous equations (\ref{Eq:waveVector}), (\ref{Eq: wn_kn}), (\ref{Eq: w_IAWn}) and iterative method, assuming the sound velocity of IAW $c_s=0.0208v_{te}$ obtained from Eq. (\ref{Eq:Dispersion_C}) keeps constant, one can obtain the precise wave numbers and frequencies of the pump light and scattered lights: 
\begin{equation}
\begin{aligned}
\label{Eq: k_w_theoretical}
&[k_0, k_1, k_2, k_3, k_4, k_5, ...]=\\
&[0.8367, -0.8326, 0.8285, -0.8204, -0.8163, ...]*\omega_0/c, \\
&[\omega_0, \omega_1, \omega_2, \omega_3, \omega_4, \omega_5, ...]=\\
&[1.0000, 0.9966, 0.9932, 0.9898, 0.9864, 0.9830, ...]*\omega_0.
\end{aligned}
\end{equation}
The results through iterative method are consistent to the approximate method above.

\begin{figure}[!tp]
 	\includegraphics[width=1\columnwidth]{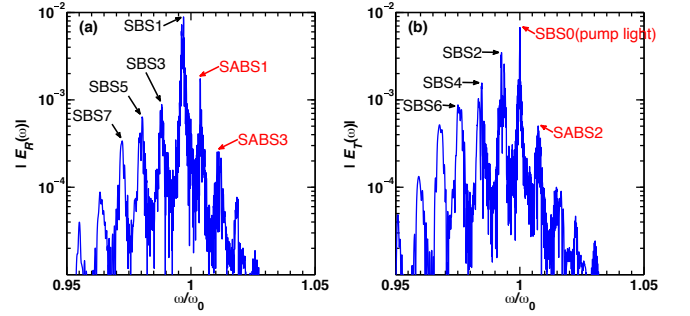}
 	
 	\caption{\label{Fig:ER_ET_w}(Color online) The frequency spectra of (a) reflective EMW electric field $E_R$ at the left boundary (incident boundary) and (b) transmitting EMW electric field $E_T$ at the right boundary (transmitting boundary). The parameters are $n_e=0.3n_c, T_e=5keV, I_0=1\times10^{16}W/cm^2$ in C plasmas as the same as Fig. \ref{Fig:w_k} and the spectra analysis time is the total simulation time $[0, 1\times10^5\omega_0^{-1}]$. }
 \end{figure}

 Figure \ref{Fig:ER_ET_w} shows the frequency spectra of the reflective EMW and the transmitting EMW. The frequencies of the Stokes scattered lights (denoted as SBS1, SBS2, SBS3, SBS4, SBS5, ... in Fig. \ref{Fig:ER_ET_w}) are:
\begin{equation}
\begin{aligned}
&[\omega'_0, \omega'_1, \omega'_2, \omega'_3, \omega'_4, \omega'_5, ...]=\\
&[1.0000, 0.9970, 0.9926, 0.9882, 0.9847, 0.9804, ...]*\omega_0,
\end{aligned}
\end{equation}
where the symbol prime represents the simulation values obtained from Fig. \ref{Fig:ER_ET_w}. The simulation results are close to the theoretical values in Eq. (\ref{Eq: k_w_theoretical}).
 In the same way, the frequency of anti-Stokes scattered light $\omega_{-1}$ in the SABS1 is: $\omega_{-1}=\omega_0+\omega_{IAW2}=\omega_0+3.4\times10^{-3}\omega_0=1.0034\omega_0$. And the frequency of anti-Stokes scattered light $\omega_{-2}$ in the SABS2 is: $\omega_{-2}=\omega_{-1}+\omega_{IAW1}=1.0034\omega_0+0.0034\omega_0=1.0068\omega_0$. 
 Thus, $\omega_{-n}=\omega_0+n*\omega_{IAW1}=(1+3.4\times10^{-3}*n)\omega_0$, i.e., 
 \begin{equation}
 \begin{aligned}
 &[\omega_{-1}, \omega_{-2},\omega_{-3}, \omega_{-4}, ...]=\\
 &[1.0034, 1.0068, 1.0102, 1.0136, ...]*\omega_0.
 \end{aligned}
 \end{equation}
 The simulation results from Fig. \ref{Fig:ER_ET_w} are:
 \begin{equation}
 \begin{aligned}
 &[\omega'_{-1}, \omega'_{-2}, \omega'_{-3}, \omega'_{-4}]=\\
 &[1.004, 1.007, 1.011, 1.015, ...]*\omega_0,
 \end{aligned}
 \end{equation}
which are consistent to the theoretical analyses above.

\begin{figure}[!tp]
	\includegraphics[width=1\columnwidth]{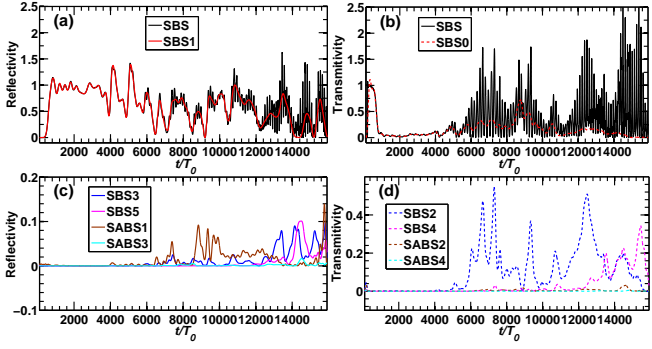}
	
	\caption{\label{Fig:Reflectivity}(Color online) The reflectivity at the left boundary and transmitivity (or scattering rate) at the right boundary of SBS cascade. (a) The reflectivity of the total SBS (black solid line) in the frequency scope $[0.9\omega_0, 0.999\omega_0]$ and the first stage stimulated Brillouin scattering (SBS1) in the frequency scope [0.993$\omega_0$, 0.999$\omega_0$]. (b) The total transmitivity (plus scattering) at the right boundary (transmitting boundary) of SBS (black solid line, $[0.9\omega_0, 1.2\omega_0]$) and the transmitivity of pump light denoted as SBS0 ($[0.997\omega_0, 1.004\omega_0]$). (c) The reflectivity of SBS3 ($[0.984\omega_0, 0.993\omega_0]$), SBS5 ($[0.976\omega_0, 0.984\omega_0]$), SABS1 ($[1.0\omega_0, 1.008\omega_0]$) and SABS3 ($[1.008\omega_0, 1.016\omega_0]$) at the incident boundary. (d) The scattering rate of SBS2 ($[0.989\omega_0, 0.997\omega_0]$), SBS4 ($[0.981\omega_0, 0.989\omega_0]$), SABS2 ($[1.004\omega_0, 1.012\omega_0]$) and SABS4 ($[1.012\omega_0, 1.020\omega_0]$) at the transmitting boundary.}
\end{figure}
\begin{figure*}
	\includegraphics[width=2\columnwidth]{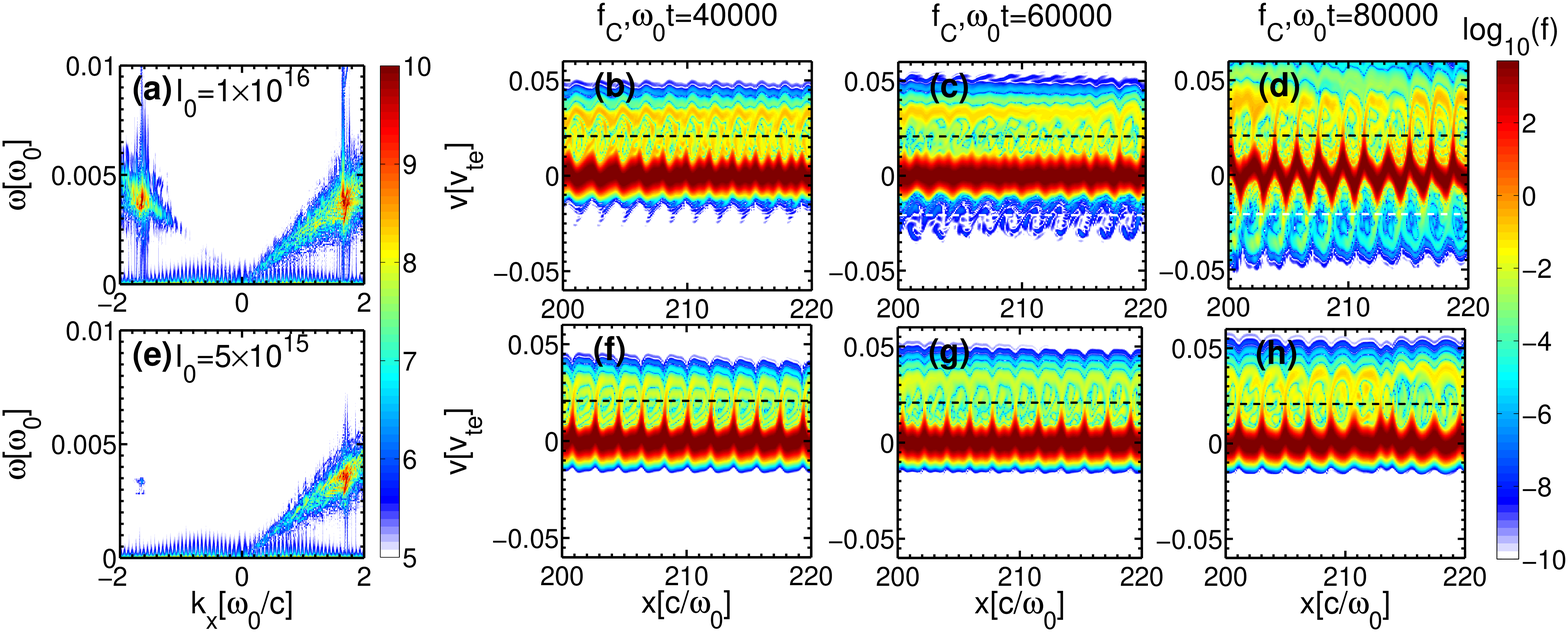}
	\caption{\label{Fig:PhasePicture}(Color online) The dispersion relations of IAWs in the condition of (a) $I_0=1\times10^{16}W/cm^2$ and (e) $I_0=5\times10^{15}W/cm^2$. The C ions distributions at the time of (b) $\omega_0t=20000$, (c) $\omega_0t=60000$, (d) $\omega_0t=80000$ in the condition of $I_0=1\times10^{16}W/cm^2$ and (f) $\omega_0t=20000$, (g) $\omega_0t=60000$, (h) $\omega_0t=80000$ in the condition of $I_0=5\times10^{15}W/cm^2$. The black dashed lines are the phase velocities of IAW1, and the white dashed lines are the phase velocities of IAW2. }
\end{figure*}
Figure \ref{Fig:Reflectivity} gives a clear demonstration of the evolution of Stokes scattering and anti-Stokes scattering in SBS cascade. We can see that the burst behaviors of the total reflectivity and transmitivity are due to the occurrence  of SBS cascade and SABS. Before $t\simeq6000T_0$, only the SBS1 ($\omega_{1}\in[0.993\omega_0, 0.999\omega_0]$) occurs, thus the total SBS reflectivity ($\omega\in[0.9\omega_0, 0.999\omega_0]$) is totally from the SBS1 as shown in Fig. \ref{Fig:Reflectivity}(a), and total SBS transmitivity ($\omega\in[0.9\omega_0, 1.2\omega_0]$) is totally from the transmitting of the pump light (SBS0, $\omega_0\in[0.997\omega_0, 1.004\omega_0]$) as shown in Fig. \ref{Fig:Reflectivity}(b). After $6000T_0$, the SBS2 ($\omega_2\in[0.989\omega_0, 0.997\omega_0]$) will occur and the strength of SBS2 is much larger than other higher stage SBS or SABS as shown in Fig. \ref{Fig:Reflectivity}(d). As a result, the strong IAW2 fluctuations will be produced by SBS2. Once the IAW2 is generated, the strong pump light will couple with IAW2 to produce a novel scattered light with a frequency larger than the pump light frequency, which is called stimulated anti-Stokes Brillouin scattering (SABS). As shown in Fig. \ref{Fig:Reflectivity}(c), the SABS1 ($\omega_{-1}\in[1.0\omega_0, 1.008\omega_0]$)) peak occurs during the time $[8000T_0, 10000T_0]$, which is even larger than SBS3. After $t\simeq12000T_0$, the SBS3, SBS4, SBS5 will develop and will compete with each other. However, the second stage SABS (SABS2, $\omega_{-2}\in[1.004\omega_0, 1.012\omega_0]$) is very weak, as a result, SABS3 and SABS4 are weaker than SABS2 and their effects on reflectivity and transmitivity can be neglected. 

Figure \ref{Fig:PhasePicture} gives the comparison of the cases in the condition of $I_0=1\times10^{16}W/cm^2$ and $I_0=5\times10^{15}W/cm^2$. As we can see, the dispersion relation of longitudinal electrostatic waves in the condition of $I_0=1\times10^{16}W/cm^2$ (Fig. \ref{Fig:PhasePicture}(a)) demonstrates two branches of IAWs, while that in the condition of $I_0=5\times10^{15}W/cm^2$ (Fig.  \ref{Fig:PhasePicture}(e)) demonstrates only one branch of IAW. This illustrates that only when the pump light intensity reaches a certain value, such as $I_0=1\times10^{16}W/cm^2$, can the SBS cascade and SABS occur. The broaden of frequency is due to the nonlinear frequency shift of IAWs \cite{Feng_2016PRE, Berger_2013POP}. Figs. \ref{Fig:PhasePicture}(b)-\ref{Fig:PhasePicture}(d) give a snapshot of the C ions distributions evolution with time in the condition of $I_0=1\times10^{16}W/cm^2$. At the time $t=40000\omega_0^{-1}=6366T_0$ (Fig. \ref{Fig:PhasePicture}(b)), the IAW1 generated by SBS1 has reached a large amplitude, thus the trapping width $\Delta v_{tr}=4\sqrt{q_i\phi/m_i}$ (i represents C ions, $q_i, m_i$ are the charge and mass of C ions, and $\phi$ is the electric potential of the IAW) will reach a large value. At the same time, the IAW2 generated by SBS2 starts to develop as shown in Fig. \ref{Fig:Reflectivity}(d) after $t\simeq6000T_0$. Figs. \ref{Fig:PhasePicture}(b)-\ref{Fig:PhasePicture}(d) show the development of IAW2. With time increasing, the ions trapping width generated by IAW2 increases obviously, this illustrates that the IAW2 amplitude increases obviously. At the time $t=60000\omega_0^{-1}=9549T_0$, the IAW2 reaches a certain amplitude as shown in Fig. \ref{Fig:PhasePicture}(c), thus, the strong pump light can couple with IAW2 to produce a higher frequency scattered light called stimulated anti-Stokes Brillouin scattering (SABS1, as shown in Fig. \ref{Fig:Reflectivity}(c)). After $t\simeq12000T_0$,  the SBS3, SBS4, SBS5 will develop as shown in Fig. \ref{Fig:Reflectivity}, thus, the positive propagating IAWs as the same as IAW1 will be generated by SBS3 and SBS5, while the negative propagating IAWs as the same as IAW2 will be generated by SBS4. Therefore, the IAW1 and IAW2 amplitudes will further increase as shown in Fig. \ref{Fig:PhasePicture}(d) at the time $t=8000\omega_0^{-1}=12732T_0$. However, if the pump light intensity decreases, such as $I_0=5\times10^{15}W/cm^2$, the IAW2 will not be generated, thus the SBS cascade and SABS will not occur. Figs. \ref{Fig:PhasePicture}(f)-\ref{Fig:PhasePicture}(h) demonstrate that the C ions will be trapped only by the positive propagating IAW1, but not the negative propagating IAW2 in the condition of $I_0=5\times10^{15}W/cm^2$. Thus, we believe that if the pump light intensity is lower than $5\times10^{15}W/cm^2$, the SBS cascade and the SABS can nearly not occur in our simulation system.

\begin{figure}[!tp]
 	\includegraphics[width=1\columnwidth]{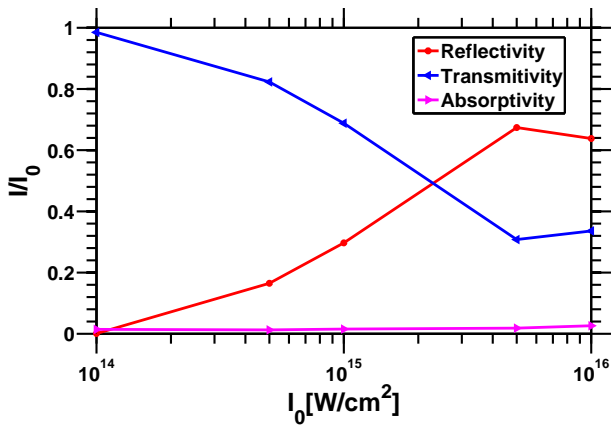}
 	
 	\caption{\label{Fig:R_T_A}(Color online) The average reflectivity, transmitivity and absorptivity of SBS vary with the pump laser intensity.}
 \end{figure}

To research the effect of SBS cascade and SABS on the reflectivity and transmitivity, the other conditions are invariant, and we change the pump light intensity, which is common intensity scale in ICF experiment. As shown in Fig. \ref{Fig:R_T_A}, if the pump light intensity is not larger than $5\times10^{15}W/cm^2$, only the first stage stimulate Brillouin scattering (SBS1) will occur as discussed above. As the pump light intensity increases from $1\times10^{14}W/cm^2$ to $5\times10^{15}W/cm^2$, the reflectivity of SBS will increase obviously and the transimitivity will decrease. As we can see, the absorptivity in the SBS process is very low (not larger than $3\%$), thus the reflectivity and the transmitivity is inversely correlated. To our surprise, when the pump intensity reaches $1\times10^{16}W/cm^2$, the total reflectivity will decrease and be lower than that in the condition of $I_0=5\times10^{15}W/cm^2$. This interesting phenomenon is due to the SBS cascade and the SABS discussed above. Although the SBS1 is the strongest scattering and dominates in the SBS cascade and SABS, a part energy of scattered light of SBS1 will transfer to the other stage SBS and SABS, especially SBS2. As shown in Fig. \ref{Fig:Reflectivity}, the SBS2 is much stronger than other higher stage SBS and SABS, thus SBS2 will obtain the most energy from SBS1, and the scattered light of SBS2 will transmit from the right boundary. Therefore, SBS2 will increase the transmitivity and decrease the reflectivity. However, there exist SABS and other higher stage SBS, such as SABS1, SBS3 and SBS5, of which the scattered lights will propagate along $-x$ direction (the pump light is assumed propagating along $+x$ direction), and will reflect from the left boundary. Thus, these scattering will increase the reflectivity and decrease the transmitivity. The two inverse effects on the reflectivity or transmitivity will compete with each other. Because the effect of SBS2 on the reflectivity or the transmitivity is the strongest in the higher stage SBS and SABS, the total reflectivity will decrease and 
the total transmitivity will increase.

In conclusions, a detail research of anti-Stokes scattering and Stokes scattering in SBS cascade in C plasmas has been carried out. An insight into the mechanism of stimulated anti-Stokes Brillouin scattering (SABS) in SBS cascade is given. The evolution of SABS and high stage SBS have been demonstrated for the first time. When the SBS cascade and SABS occur, the reflectivity and transmitivity will appear a burst behavior. Since the effect of SBS2 is the strongest in the higher stage SBS and SABS, the reflectivity will decrease and the transmitivity will increase. These  interesting results give a advance and of important significance to the field of high-intensity laser particle interaction in ICF.

We are pleased to acknowledge useful discussions with K. Q. Pan, T. W. Huang, J. X. Gong and B. Qiao. This research was supported by the National Natural Science Foundation of China (Grant Nos. 11575035, 11475030 and 11435011) and National Basic Research
Program of China (Grant No. 2013CB834101).

\bibliography{SBS}

\begin{thebibliography}{31}%
\makeatletter
\providecommand \@ifxundefined [1]{%
 \@ifx{#1\undefined}
}%
\providecommand \@ifnum [1]{%
 \ifnum #1\expandafter \@firstoftwo
 \else \expandafter \@secondoftwo
 \fi
}%
\providecommand \@ifx [1]{%
 \ifx #1\expandafter \@firstoftwo
 \else \expandafter \@secondoftwo
 \fi
}%
\providecommand \natexlab [1]{#1}%
\providecommand \enquote  [1]{``#1''}%
\providecommand \bibnamefont  [1]{#1}%
\providecommand \bibfnamefont [1]{#1}%
\providecommand \citenamefont [1]{#1}%
\providecommand \href@noop [0]{\@secondoftwo}%
\providecommand \href [0]{\begingroup \@sanitize@url \@href}%
\providecommand \@href[1]{\@@startlink{#1}\@@href}%
\providecommand \@@href[1]{\endgroup#1\@@endlink}%
\providecommand \@sanitize@url [0]{\catcode `\\12\catcode `\$12\catcode
  `\&12\catcode `\#12\catcode `\^12\catcode `\_12\catcode `\%12\relax}%
\providecommand \@@startlink[1]{}%
\providecommand \@@endlink[0]{}%
\providecommand \url  [0]{\begingroup\@sanitize@url \@url }%
\providecommand \@url [1]{\endgroup\@href {#1}{\urlprefix }}%
\providecommand \urlprefix  [0]{URL }%
\providecommand \Eprint [0]{\href }%
\providecommand \doibase [0]{http://dx.doi.org/}%
\providecommand \selectlanguage [0]{\@gobble}%
\providecommand \bibinfo  [0]{\@secondoftwo}%
\providecommand \bibfield  [0]{\@secondoftwo}%
\providecommand \translation [1]{[#1]}%
\providecommand \BibitemOpen [0]{}%
\providecommand \bibitemStop [0]{}%
\providecommand \bibitemNoStop [0]{.\EOS\space}%
\providecommand \EOS [0]{\spacefactor3000\relax}%
\providecommand \BibitemShut  [1]{\csname bibitem#1\endcsname}%
\let\auto@bib@innerbib\@empty
\bibitem [{\citenamefont {He}\ \emph {et~al.}(2016)\citenamefont {He},
  \citenamefont {Li}, \citenamefont {Fan}, \citenamefont {Wang}, \citenamefont
  {Liu}, \citenamefont {Lan}, \citenamefont {Wu},\ and\ \citenamefont
  {Ye}}]{He_2016POP}%
  \BibitemOpen
  \bibfield  {author} {\bibinfo {author} {\bibfnamefont {X.~T.}\ \bibnamefont
  {He}}, \bibinfo {author} {\bibfnamefont {J.~W.}\ \bibnamefont {Li}}, \bibinfo
  {author} {\bibfnamefont {Z.~F.}\ \bibnamefont {Fan}}, \bibinfo {author}
  {\bibfnamefont {L.~F.}\ \bibnamefont {Wang}}, \bibinfo {author}
  {\bibfnamefont {J.}~\bibnamefont {Liu}}, \bibinfo {author} {\bibfnamefont
  {K.}~\bibnamefont {Lan}}, \bibinfo {author} {\bibfnamefont {J.~F.}\
  \bibnamefont {Wu}}, \ and\ \bibinfo {author} {\bibfnamefont {W.~H.}\
  \bibnamefont {Ye}},\ }\href {\doibase 10.1063/1.4960973} {\bibfield
  {journal} {\bibinfo  {journal} {Physics of Plasmas}\ }\textbf {\bibinfo
  {volume} {23}},\ \bibinfo {eid} {082706} (\bibinfo {year}
  {2016})}\BibitemShut {NoStop}%
\bibitem [{\citenamefont {Glenzer}\ \emph {et~al.}(2010)\citenamefont
  {Glenzer}, \citenamefont {MacGowan}, \citenamefont {Michel}, \citenamefont
  {Meezan}, \citenamefont {Suter}, \citenamefont {Dixit}, \citenamefont
  {Kline}, \citenamefont {Kyrala}, \citenamefont {Bradley}, \citenamefont
  {Callahan}, \citenamefont {Dewald}, \citenamefont {Divol}, \citenamefont
  {Dzenitis}, \citenamefont {Edwards}, \citenamefont {Hamza}, \citenamefont
  {Haynam}, \citenamefont {Hinkel}, \citenamefont {Kalantar}, \citenamefont
  {Kilkenny}, \citenamefont {Landen}, \citenamefont {Lindl}, \citenamefont
  {LePape}, \citenamefont {Moody}, \citenamefont {Nikroo}, \citenamefont
  {Parham}, \citenamefont {Schneider}, \citenamefont {Town}, \citenamefont
  {Wegner}, \citenamefont {Widmann}, \citenamefont {Whitman}, \citenamefont
  {Young}, \citenamefont {Van~Wonterghem}, \citenamefont {Atherton},\ and\
  \citenamefont {Moses}}]{Glenzer_2010Science}%
  \BibitemOpen
  \bibfield  {author} {\bibinfo {author} {\bibfnamefont {S.~H.}\ \bibnamefont
  {Glenzer}}, \bibinfo {author} {\bibfnamefont {B.~J.}\ \bibnamefont
  {MacGowan}}, \bibinfo {author} {\bibfnamefont {P.}~\bibnamefont {Michel}},
  \bibinfo {author} {\bibfnamefont {N.~B.}\ \bibnamefont {Meezan}}, \bibinfo
  {author} {\bibfnamefont {L.~J.}\ \bibnamefont {Suter}}, \bibinfo {author}
  {\bibfnamefont {S.~N.}\ \bibnamefont {Dixit}}, \bibinfo {author}
  {\bibfnamefont {J.~L.}\ \bibnamefont {Kline}}, \bibinfo {author}
  {\bibfnamefont {G.~A.}\ \bibnamefont {Kyrala}}, \bibinfo {author}
  {\bibfnamefont {D.~K.}\ \bibnamefont {Bradley}}, \bibinfo {author}
  {\bibfnamefont {D.~A.}\ \bibnamefont {Callahan}}, \bibinfo {author}
  {\bibfnamefont {E.~L.}\ \bibnamefont {Dewald}}, \bibinfo {author}
  {\bibfnamefont {L.}~\bibnamefont {Divol}}, \bibinfo {author} {\bibfnamefont
  {E.}~\bibnamefont {Dzenitis}}, \bibinfo {author} {\bibfnamefont {M.~J.}\
  \bibnamefont {Edwards}}, \bibinfo {author} {\bibfnamefont {A.~V.}\
  \bibnamefont {Hamza}}, \bibinfo {author} {\bibfnamefont {C.~A.}\ \bibnamefont
  {Haynam}}, \bibinfo {author} {\bibfnamefont {D.~E.}\ \bibnamefont {Hinkel}},
  \bibinfo {author} {\bibfnamefont {D.~H.}\ \bibnamefont {Kalantar}}, \bibinfo
  {author} {\bibfnamefont {J.~D.}\ \bibnamefont {Kilkenny}}, \bibinfo {author}
  {\bibfnamefont {O.~L.}\ \bibnamefont {Landen}}, \bibinfo {author}
  {\bibfnamefont {J.~D.}\ \bibnamefont {Lindl}}, \bibinfo {author}
  {\bibfnamefont {S.}~\bibnamefont {LePape}}, \bibinfo {author} {\bibfnamefont
  {J.~D.}\ \bibnamefont {Moody}}, \bibinfo {author} {\bibfnamefont
  {A.}~\bibnamefont {Nikroo}}, \bibinfo {author} {\bibfnamefont
  {T.}~\bibnamefont {Parham}}, \bibinfo {author} {\bibfnamefont {M.~B.}\
  \bibnamefont {Schneider}}, \bibinfo {author} {\bibfnamefont {R.~P.~J.}\
  \bibnamefont {Town}}, \bibinfo {author} {\bibfnamefont {P.}~\bibnamefont
  {Wegner}}, \bibinfo {author} {\bibfnamefont {K.}~\bibnamefont {Widmann}},
  \bibinfo {author} {\bibfnamefont {P.}~\bibnamefont {Whitman}}, \bibinfo
  {author} {\bibfnamefont {B.~K.~F.}\ \bibnamefont {Young}}, \bibinfo {author}
  {\bibfnamefont {B.}~\bibnamefont {Van~Wonterghem}}, \bibinfo {author}
  {\bibfnamefont {L.~J.}\ \bibnamefont {Atherton}}, \ and\ \bibinfo {author}
  {\bibfnamefont {E.~I.}\ \bibnamefont {Moses}},\ }\href {\doibase
  10.1126/science.1185634} {\bibfield  {journal} {\bibinfo  {journal}
  {Science}\ }\textbf {\bibinfo {volume} {327}},\ \bibinfo {pages} {228}
  (\bibinfo {year} {2010})}\BibitemShut {NoStop}%
\bibitem [{\citenamefont {Glenzer}\ \emph {et~al.}(2007)\citenamefont
  {Glenzer}, \citenamefont {Froula}, \citenamefont {Divol}, \citenamefont
  {Dorr}, \citenamefont {Berger}, \citenamefont {Dixit}, \citenamefont
  {Hammel}, \citenamefont {Haynam}, \citenamefont {Hittinger}, \citenamefont
  {Holder}, \citenamefont {Jones}, \citenamefont {Kalantar}, \citenamefont
  {Landen}, \citenamefont {Langdon}, \citenamefont {Langer}, \citenamefont
  {MacGowan}, \citenamefont {Mackinnon}, \citenamefont {Meezan}, \citenamefont
  {Moses}, \citenamefont {Niemann}, \citenamefont {Still}, \citenamefont
  {Suter}, \citenamefont {Wallace}, \citenamefont {Williams},\ and\
  \citenamefont {Young}}]{Glenzer_2007Nature}%
  \BibitemOpen
  \bibfield  {author} {\bibinfo {author} {\bibfnamefont {S.~H.}\ \bibnamefont
  {Glenzer}}, \bibinfo {author} {\bibfnamefont {D.~H.}\ \bibnamefont {Froula}},
  \bibinfo {author} {\bibfnamefont {L.}~\bibnamefont {Divol}}, \bibinfo
  {author} {\bibfnamefont {M.}~\bibnamefont {Dorr}}, \bibinfo {author}
  {\bibfnamefont {R.~L.}\ \bibnamefont {Berger}}, \bibinfo {author}
  {\bibfnamefont {S.}~\bibnamefont {Dixit}}, \bibinfo {author} {\bibfnamefont
  {B.~A.}\ \bibnamefont {Hammel}}, \bibinfo {author} {\bibfnamefont
  {C.}~\bibnamefont {Haynam}}, \bibinfo {author} {\bibfnamefont {J.~A.}\
  \bibnamefont {Hittinger}}, \bibinfo {author} {\bibfnamefont {J.~P.}\
  \bibnamefont {Holder}}, \bibinfo {author} {\bibfnamefont {O.~S.}\
  \bibnamefont {Jones}}, \bibinfo {author} {\bibfnamefont {D.~H.}\ \bibnamefont
  {Kalantar}}, \bibinfo {author} {\bibfnamefont {O.~L.}\ \bibnamefont
  {Landen}}, \bibinfo {author} {\bibfnamefont {A.~B.}\ \bibnamefont {Langdon}},
  \bibinfo {author} {\bibfnamefont {S.}~\bibnamefont {Langer}}, \bibinfo
  {author} {\bibfnamefont {B.~J.}\ \bibnamefont {MacGowan}}, \bibinfo {author}
  {\bibfnamefont {A.~J.}\ \bibnamefont {Mackinnon}}, \bibinfo {author}
  {\bibfnamefont {N.}~\bibnamefont {Meezan}}, \bibinfo {author} {\bibfnamefont
  {E.~I.}\ \bibnamefont {Moses}}, \bibinfo {author} {\bibfnamefont
  {C.}~\bibnamefont {Niemann}}, \bibinfo {author} {\bibfnamefont {C.~H.}\
  \bibnamefont {Still}}, \bibinfo {author} {\bibfnamefont {L.~J.}\ \bibnamefont
  {Suter}}, \bibinfo {author} {\bibfnamefont {R.~J.}\ \bibnamefont {Wallace}},
  \bibinfo {author} {\bibfnamefont {E.~A.}\ \bibnamefont {Williams}}, \ and\
  \bibinfo {author} {\bibfnamefont {B.~K.~F.}\ \bibnamefont {Young}},\ }\href
  {http://dx.doi.org/10.1038/nphys709} {\bibfield  {journal} {\bibinfo
  {journal} {Nat. Phys.}\ }\textbf {\bibinfo {volume} {3}},\ \bibinfo {pages}
  {716} (\bibinfo {year} {2007})}\BibitemShut {NoStop}%
\bibitem [{\citenamefont {Liu}\ \emph {et~al.}(2009{\natexlab{a}})\citenamefont
  {Liu}, \citenamefont {He}, \citenamefont {Zheng},\ and\ \citenamefont
  {Wang}}]{Liu_2009POP_1}%
  \BibitemOpen
  \bibfield  {author} {\bibinfo {author} {\bibfnamefont {Z.~J.}\ \bibnamefont
  {Liu}}, \bibinfo {author} {\bibfnamefont {X.~T.}\ \bibnamefont {He}},
  \bibinfo {author} {\bibfnamefont {C.~Y.}\ \bibnamefont {Zheng}}, \ and\
  \bibinfo {author} {\bibfnamefont {Y.~G.}\ \bibnamefont {Wang}},\ }\href
  {\doibase 10.1063/1.3227647} {\bibfield  {journal} {\bibinfo  {journal}
  {Physics of Plasmas}\ }\textbf {\bibinfo {volume} {16}},\ \bibinfo {eid}
  {093108} (\bibinfo {year} {2009}{\natexlab{a}})}\BibitemShut {NoStop}%
\bibitem [{\citenamefont {Weber}\ \emph
  {et~al.}(2005{\natexlab{a}})\citenamefont {Weber}, \citenamefont {Riconda},\
  and\ \citenamefont {Tikhonchuk}}]{Weber_2005PRL}%
  \BibitemOpen
  \bibfield  {author} {\bibinfo {author} {\bibfnamefont {S.}~\bibnamefont
  {Weber}}, \bibinfo {author} {\bibfnamefont {C.}~\bibnamefont {Riconda}}, \
  and\ \bibinfo {author} {\bibfnamefont {V.~T.}\ \bibnamefont {Tikhonchuk}},\
  }\href {\doibase 10.1103/PhysRevLett.94.055005} {\bibfield  {journal}
  {\bibinfo  {journal} {Phys. Rev. Lett.}\ }\textbf {\bibinfo {volume} {94}},\
  \bibinfo {pages} {055005} (\bibinfo {year} {2005}{\natexlab{a}})}\BibitemShut
  {NoStop}%
\bibitem [{\citenamefont {Weber}\ \emph
  {et~al.}(2005{\natexlab{b}})\citenamefont {Weber}, \citenamefont {Riconda},\
  and\ \citenamefont {Tikhonchuk}}]{Weber_2005POP}%
  \BibitemOpen
  \bibfield  {author} {\bibinfo {author} {\bibfnamefont {S.}~\bibnamefont
  {Weber}}, \bibinfo {author} {\bibfnamefont {C.}~\bibnamefont {Riconda}}, \
  and\ \bibinfo {author} {\bibfnamefont {V.~T.}\ \bibnamefont {Tikhonchuk}},\
  }\href {\doibase 10.1063/1.1862246} {\bibfield  {journal} {\bibinfo
  {journal} {Physics of Plasmas}\ }\textbf {\bibinfo {volume} {12}},\ \bibinfo
  {pages} {043101} (\bibinfo {year} {2005}{\natexlab{b}})}\BibitemShut
  {NoStop}%
\bibitem [{\citenamefont {Weber}\ \emph
  {et~al.}(2005{\natexlab{c}})\citenamefont {Weber}, \citenamefont {Lontano},
  \citenamefont {Passoni}, \citenamefont {Riconda},\ and\ \citenamefont
  {Tikhonchuk}}]{Weber_2005POP_1}%
  \BibitemOpen
  \bibfield  {author} {\bibinfo {author} {\bibfnamefont {S.}~\bibnamefont
  {Weber}}, \bibinfo {author} {\bibfnamefont {M.}~\bibnamefont {Lontano}},
  \bibinfo {author} {\bibfnamefont {M.}~\bibnamefont {Passoni}}, \bibinfo
  {author} {\bibfnamefont {C.}~\bibnamefont {Riconda}}, \ and\ \bibinfo
  {author} {\bibfnamefont {V.~T.}\ \bibnamefont {Tikhonchuk}},\ }\href
  {\doibase 10.1063/1.2136354} {\bibfield  {journal} {\bibinfo  {journal}
  {Physics of Plasmas}\ }\textbf {\bibinfo {volume} {12}},\ \bibinfo {pages}
  {112107} (\bibinfo {year} {2005}{\natexlab{c}})}\BibitemShut {NoStop}%
\bibitem [{\citenamefont {Froula}\ \emph {et~al.}(2002)\citenamefont {Froula},
  \citenamefont {Divol},\ and\ \citenamefont {Glenzer}}]{Froula_2002PRL}%
  \BibitemOpen
  \bibfield  {author} {\bibinfo {author} {\bibfnamefont {D.~H.}\ \bibnamefont
  {Froula}}, \bibinfo {author} {\bibfnamefont {L.}~\bibnamefont {Divol}}, \
  and\ \bibinfo {author} {\bibfnamefont {S.~H.}\ \bibnamefont {Glenzer}},\
  }\href {\doibase 10.1103/PhysRevLett.88.105003} {\bibfield  {journal}
  {\bibinfo  {journal} {Phys. Rev. Lett.}\ }\textbf {\bibinfo {volume} {88}},\
  \bibinfo {pages} {105003} (\bibinfo {year} {2002})}\BibitemShut {NoStop}%
\bibitem [{\citenamefont {Giacone}\ and\ \citenamefont
  {Vu}(1998)}]{Giacone_1998POP}%
  \BibitemOpen
  \bibfield  {author} {\bibinfo {author} {\bibfnamefont {R.~E.}\ \bibnamefont
  {Giacone}}\ and\ \bibinfo {author} {\bibfnamefont {H.~X.}\ \bibnamefont
  {Vu}},\ }\href {\doibase 10.1063/1.872803} {\bibfield  {journal} {\bibinfo
  {journal} {Physics of Plasmas}\ }\textbf {\bibinfo {volume} {5}},\ \bibinfo
  {pages} {1455} (\bibinfo {year} {1998})}\BibitemShut {NoStop}%
\bibitem [{\citenamefont {Vu}\ \emph {et~al.}(2001)\citenamefont {Vu},
  \citenamefont {DuBois},\ and\ \citenamefont {Bezzerides}}]{Vu_2001PRL}%
  \BibitemOpen
  \bibfield  {author} {\bibinfo {author} {\bibfnamefont {H.~X.}\ \bibnamefont
  {Vu}}, \bibinfo {author} {\bibfnamefont {D.~F.}\ \bibnamefont {DuBois}}, \
  and\ \bibinfo {author} {\bibfnamefont {B.}~\bibnamefont {Bezzerides}},\
  }\href {\doibase 10.1103/PhysRevLett.86.4306} {\bibfield  {journal} {\bibinfo
   {journal} {Phys. Rev. Lett.}\ }\textbf {\bibinfo {volume} {86}},\ \bibinfo
  {pages} {4306} (\bibinfo {year} {2001})}\BibitemShut {NoStop}%
\bibitem [{\citenamefont {Albright}\ \emph {et~al.}(2016)\citenamefont
  {Albright}, \citenamefont {Yin}, \citenamefont {Bowers},\ and\ \citenamefont
  {Bergen}}]{Albright_2016POP}%
  \BibitemOpen
  \bibfield  {author} {\bibinfo {author} {\bibfnamefont {B.~J.}\ \bibnamefont
  {Albright}}, \bibinfo {author} {\bibfnamefont {L.}~\bibnamefont {Yin}},
  \bibinfo {author} {\bibfnamefont {K.~J.}\ \bibnamefont {Bowers}}, \ and\
  \bibinfo {author} {\bibfnamefont {B.}~\bibnamefont {Bergen}},\ }\href
  {\doibase 10.1063/1.4943102} {\bibfield  {journal} {\bibinfo  {journal}
  {Physics of Plasmas}\ }\textbf {\bibinfo {volume} {23}},\ \bibinfo {pages}
  {032703} (\bibinfo {year} {2016})}\BibitemShut {NoStop}%
\bibitem [{\citenamefont {Cohen}\ \emph {et~al.}(1997)\citenamefont {Cohen},
  \citenamefont {Lasinski}, \citenamefont {Langdon},\ and\ \citenamefont
  {Williams}}]{Bruce_1997POP}%
  \BibitemOpen
  \bibfield  {author} {\bibinfo {author} {\bibfnamefont {B.~I.}\ \bibnamefont
  {Cohen}}, \bibinfo {author} {\bibfnamefont {B.~F.}\ \bibnamefont {Lasinski}},
  \bibinfo {author} {\bibfnamefont {A.~B.}\ \bibnamefont {Langdon}}, \ and\
  \bibinfo {author} {\bibfnamefont {E.~A.}\ \bibnamefont {Williams}},\ }\href
  {\doibase 10.1063/1.872187} {\bibfield  {journal} {\bibinfo  {journal}
  {Physics of Plasmas}\ }\textbf {\bibinfo {volume} {4}},\ \bibinfo {pages}
  {956} (\bibinfo {year} {1997})}\BibitemShut {NoStop}%
\bibitem [{\citenamefont {Rozmus}\ \emph {et~al.}(1992)\citenamefont {Rozmus},
  \citenamefont {Casanova}, \citenamefont {Pesme}, \citenamefont {Heron},\ and\
  \citenamefont {Adam}}]{Rozmus_1992POP}%
  \BibitemOpen
  \bibfield  {author} {\bibinfo {author} {\bibfnamefont {W.}~\bibnamefont
  {Rozmus}}, \bibinfo {author} {\bibfnamefont {M.}~\bibnamefont {Casanova}},
  \bibinfo {author} {\bibfnamefont {D.}~\bibnamefont {Pesme}}, \bibinfo
  {author} {\bibfnamefont {A.}~\bibnamefont {Heron}}, \ and\ \bibinfo {author}
  {\bibfnamefont {J.}~\bibnamefont {Adam}},\ }\href {\doibase 10.1063/1.860256}
  {\bibfield  {journal} {\bibinfo  {journal} {Physics of Fluids B: Plasma
  Physics}\ }\textbf {\bibinfo {volume} {4}},\ \bibinfo {pages} {576} (\bibinfo
  {year} {1992})}\BibitemShut {NoStop}%
\bibitem [{\citenamefont {Rambo}\ \emph {et~al.}(1997)\citenamefont {Rambo},
  \citenamefont {Wilks},\ and\ \citenamefont {Kruer}}]{Rambo_1997PRL}%
  \BibitemOpen
  \bibfield  {author} {\bibinfo {author} {\bibfnamefont {P.~W.}\ \bibnamefont
  {Rambo}}, \bibinfo {author} {\bibfnamefont {S.~C.}\ \bibnamefont {Wilks}}, \
  and\ \bibinfo {author} {\bibfnamefont {W.~L.}\ \bibnamefont {Kruer}},\ }\href
  {\doibase 10.1103/PhysRevLett.79.83} {\bibfield  {journal} {\bibinfo
  {journal} {Phys. Rev. Lett.}\ }\textbf {\bibinfo {volume} {79}},\ \bibinfo
  {pages} {83} (\bibinfo {year} {1997})}\BibitemShut {NoStop}%
\bibitem [{\citenamefont {Pawley}\ \emph {et~al.}(1982)\citenamefont {Pawley},
  \citenamefont {Huey},\ and\ \citenamefont {Luhmann}}]{Pawley_1982PRL}%
  \BibitemOpen
  \bibfield  {author} {\bibinfo {author} {\bibfnamefont {C.~J.}\ \bibnamefont
  {Pawley}}, \bibinfo {author} {\bibfnamefont {H.~E.}\ \bibnamefont {Huey}}, \
  and\ \bibinfo {author} {\bibfnamefont {N.~C.}\ \bibnamefont {Luhmann}},\
  }\href {\doibase 10.1103/PhysRevLett.49.877} {\bibfield  {journal} {\bibinfo
  {journal} {Phys. Rev. Lett.}\ }\textbf {\bibinfo {volume} {49}},\ \bibinfo
  {pages} {877} (\bibinfo {year} {1982})}\BibitemShut {NoStop}%
\bibitem [{\citenamefont {Speziale}\ \emph {et~al.}(1980)\citenamefont
  {Speziale}, \citenamefont {McGrath},\ and\ \citenamefont
  {Berger}}]{Speziale_1980POF}%
  \BibitemOpen
  \bibfield  {author} {\bibinfo {author} {\bibfnamefont {T.}~\bibnamefont
  {Speziale}}, \bibinfo {author} {\bibfnamefont {J.~F.}\ \bibnamefont
  {McGrath}}, \ and\ \bibinfo {author} {\bibfnamefont {R.~L.}\ \bibnamefont
  {Berger}},\ }\href {\doibase 10.1063/1.863127} {\bibfield  {journal}
  {\bibinfo  {journal} {The Physics of Fluids}\ }\textbf {\bibinfo {volume}
  {23}},\ \bibinfo {pages} {1275} (\bibinfo {year} {1980})}\BibitemShut
  {NoStop}%
\bibitem [{\citenamefont {Zhan-Jun}\ \emph {et~al.}(2012)\citenamefont
  {Zhan-Jun}, \citenamefont {Xian-Tu}, \citenamefont {Chun-Yang},\ and\
  \citenamefont {Yu-Gang}}]{Liu_2012CPB}%
  \BibitemOpen
  \bibfield  {author} {\bibinfo {author} {\bibfnamefont {L.}~\bibnamefont
  {Zhan-Jun}}, \bibinfo {author} {\bibfnamefont {H.}~\bibnamefont {Xian-Tu}},
  \bibinfo {author} {\bibfnamefont {Z.}~\bibnamefont {Chun-Yang}}, \ and\
  \bibinfo {author} {\bibfnamefont {W.}~\bibnamefont {Yu-Gang}},\ }\href
  {http://stacks.iop.org/1674-1056/21/i=1/a=015202} {\bibfield  {journal}
  {\bibinfo  {journal} {Chinese Physics B}\ }\textbf {\bibinfo {volume} {21}},\
  \bibinfo {pages} {015202} (\bibinfo {year} {2012})}\BibitemShut {NoStop}%
\bibitem [{\citenamefont {Turner}\ and\ \citenamefont
  {Goldman}(1981)}]{Robert_1981POF}%
  \BibitemOpen
  \bibfield  {author} {\bibinfo {author} {\bibfnamefont {R.~E.}\ \bibnamefont
  {Turner}}\ and\ \bibinfo {author} {\bibfnamefont {L.~M.}\ \bibnamefont
  {Goldman}},\ }\href {\doibase 10.1063/1.863240} {\bibfield  {journal}
  {\bibinfo  {journal} {The Physics of Fluids}\ }\textbf {\bibinfo {volume}
  {24}},\ \bibinfo {pages} {184} (\bibinfo {year} {1981})}\BibitemShut
  {NoStop}%
\bibitem [{\citenamefont {R\'{e}gnier}\ and\ \citenamefont
  {Taran}(1973)}]{Regnier_1973APL}%
  \BibitemOpen
  \bibfield  {author} {\bibinfo {author} {\bibfnamefont {P.~R.}\ \bibnamefont
  {R\'{e}gnier}}\ and\ \bibinfo {author} {\bibfnamefont {J.~P.~E.}\
  \bibnamefont {Taran}},\ }\href {\doibase 10.1063/1.1654873} {\bibfield
  {journal} {\bibinfo  {journal} {Applied Physics Letters}\ }\textbf {\bibinfo
  {volume} {23}},\ \bibinfo {pages} {240} (\bibinfo {year} {1973})}\BibitemShut
  {NoStop}%
\bibitem [{\citenamefont {Hickman}\ and\ \citenamefont
  {Bischel}(1988)}]{Hickman_1988PRA}%
  \BibitemOpen
  \bibfield  {author} {\bibinfo {author} {\bibfnamefont {A.~P.}\ \bibnamefont
  {Hickman}}\ and\ \bibinfo {author} {\bibfnamefont {W.~K.}\ \bibnamefont
  {Bischel}},\ }\href {\doibase 10.1103/PhysRevA.37.2516} {\bibfield  {journal}
  {\bibinfo  {journal} {Phys. Rev. A}\ }\textbf {\bibinfo {volume} {37}},\
  \bibinfo {pages} {2516} (\bibinfo {year} {1988})}\BibitemShut {NoStop}%
\bibitem [{\citenamefont {Manz}\ \emph {et~al.}(2004)\citenamefont {Manz},
  \citenamefont {Schwarz},\ and\ \citenamefont {Maier}}]{Manz_2004OC}%
  \BibitemOpen
  \bibfield  {author} {\bibinfo {author} {\bibfnamefont {T.}~\bibnamefont
  {Manz}}, \bibinfo {author} {\bibfnamefont {U.}~\bibnamefont {Schwarz}}, \
  and\ \bibinfo {author} {\bibfnamefont {M.}~\bibnamefont {Maier}},\ }\href
  {\doibase http://dx.doi.org/10.1016/j.optcom.2004.02.047} {\bibfield
  {journal} {\bibinfo  {journal} {Optics Communications}\ }\textbf {\bibinfo
  {volume} {235}},\ \bibinfo {pages} {201 } (\bibinfo {year}
  {2004})}\BibitemShut {NoStop}%
\bibitem [{\citenamefont {Kneipp}\ \emph {et~al.}(2000)\citenamefont {Kneipp},
  \citenamefont {Kneipp}, \citenamefont {Corio}, \citenamefont {Brown},
  \citenamefont {Shafer}, \citenamefont {Motz}, \citenamefont {Perelman},
  \citenamefont {Hanlon}, \citenamefont {Marucci}, \citenamefont
  {Dresselhaus},\ and\ \citenamefont {Dresselhaus}}]{Kneipp_2000PRL}%
  \BibitemOpen
  \bibfield  {author} {\bibinfo {author} {\bibfnamefont {K.}~\bibnamefont
  {Kneipp}}, \bibinfo {author} {\bibfnamefont {H.}~\bibnamefont {Kneipp}},
  \bibinfo {author} {\bibfnamefont {P.}~\bibnamefont {Corio}}, \bibinfo
  {author} {\bibfnamefont {S.~D.~M.}\ \bibnamefont {Brown}}, \bibinfo {author}
  {\bibfnamefont {K.}~\bibnamefont {Shafer}}, \bibinfo {author} {\bibfnamefont
  {J.}~\bibnamefont {Motz}}, \bibinfo {author} {\bibfnamefont {L.~T.}\
  \bibnamefont {Perelman}}, \bibinfo {author} {\bibfnamefont {E.~B.}\
  \bibnamefont {Hanlon}}, \bibinfo {author} {\bibfnamefont {A.}~\bibnamefont
  {Marucci}}, \bibinfo {author} {\bibfnamefont {G.}~\bibnamefont
  {Dresselhaus}}, \ and\ \bibinfo {author} {\bibfnamefont {M.~S.}\ \bibnamefont
  {Dresselhaus}},\ }\href {\doibase 10.1103/PhysRevLett.84.3470} {\bibfield
  {journal} {\bibinfo  {journal} {Phys. Rev. Lett.}\ }\textbf {\bibinfo
  {volume} {84}},\ \bibinfo {pages} {3470} (\bibinfo {year}
  {2000})}\BibitemShut {NoStop}%
\bibitem [{\citenamefont {Goldblatt}\ and\ \citenamefont
  {Hercher}(1968)}]{Goldblatt_1968PRL}%
  \BibitemOpen
  \bibfield  {author} {\bibinfo {author} {\bibfnamefont {N.}~\bibnamefont
  {Goldblatt}}\ and\ \bibinfo {author} {\bibfnamefont {M.}~\bibnamefont
  {Hercher}},\ }\href {\doibase 10.1103/PhysRevLett.20.310} {\bibfield
  {journal} {\bibinfo  {journal} {Phys. Rev. Lett.}\ }\textbf {\bibinfo
  {volume} {20}},\ \bibinfo {pages} {310} (\bibinfo {year} {1968})}\BibitemShut
  {NoStop}%
\bibitem [{\citenamefont {Shin}\ \emph {et~al.}(2013)\citenamefont {Shin},
  \citenamefont {Qiu}, \citenamefont {Jarecki}, \citenamefont {Cox},
  \citenamefont {Olsson}, \citenamefont {Starbuck}, \citenamefont {Wang},\ and\
  \citenamefont {Rakich}}]{Shin_2013NC}%
  \BibitemOpen
  \bibfield  {author} {\bibinfo {author} {\bibfnamefont {H.}~\bibnamefont
  {Shin}}, \bibinfo {author} {\bibfnamefont {W.}~\bibnamefont {Qiu}}, \bibinfo
  {author} {\bibfnamefont {R.}~\bibnamefont {Jarecki}}, \bibinfo {author}
  {\bibfnamefont {J.~A.}\ \bibnamefont {Cox}}, \bibinfo {author} {\bibfnamefont
  {R.~H.}\ \bibnamefont {Olsson}}, \bibinfo {author} {\bibfnamefont
  {A.}~\bibnamefont {Starbuck}}, \bibinfo {author} {\bibfnamefont
  {Z.}~\bibnamefont {Wang}}, \ and\ \bibinfo {author} {\bibfnamefont {P.~T.}\
  \bibnamefont {Rakich}},\ }\href {http://dx.doi.org/10.1038/ncomms2943}
  {\bibfield  {journal} {\bibinfo  {journal} {Nature Communications}\ }\textbf
  {\bibinfo {volume} {4}},\ \bibinfo {pages} {1944 EP } (\bibinfo {year}
  {2013})}\BibitemShut {NoStop}%
\bibitem [{\citenamefont {Liu}\ \emph {et~al.}(2009{\natexlab{b}})\citenamefont
  {Liu}, \citenamefont {Zhu}, \citenamefont {Cao}, \citenamefont {Zheng},
  \citenamefont {He},\ and\ \citenamefont {Wang}}]{Liu_2009POP}%
  \BibitemOpen
  \bibfield  {author} {\bibinfo {author} {\bibfnamefont {Z.~J.}\ \bibnamefont
  {Liu}}, \bibinfo {author} {\bibfnamefont {S.~P.}\ \bibnamefont {Zhu}},
  \bibinfo {author} {\bibfnamefont {L.~H.}\ \bibnamefont {Cao}}, \bibinfo
  {author} {\bibfnamefont {C.~Y.}\ \bibnamefont {Zheng}}, \bibinfo {author}
  {\bibfnamefont {X.~T.}\ \bibnamefont {He}}, \ and\ \bibinfo {author}
  {\bibfnamefont {Y.}~\bibnamefont {Wang}},\ }\href {\doibase
  10.1063/1.3258839} {\bibfield  {journal} {\bibinfo  {journal} {Physics of
  Plasmas}\ }\textbf {\bibinfo {volume} {16}},\ \bibinfo {eid} {112703}
  (\bibinfo {year} {2009}{\natexlab{b}})}\BibitemShut {NoStop}%
\bibitem [{\citenamefont {Liu}\ \emph {et~al.}(2011)\citenamefont {Liu},
  \citenamefont {Zheng}, \citenamefont {He},\ and\ \citenamefont
  {Wang}}]{Liu_2011POP}%
  \BibitemOpen
  \bibfield  {author} {\bibinfo {author} {\bibfnamefont {Z.~J.}\ \bibnamefont
  {Liu}}, \bibinfo {author} {\bibfnamefont {C.~Y.}\ \bibnamefont {Zheng}},
  \bibinfo {author} {\bibfnamefont {X.~T.}\ \bibnamefont {He}}, \ and\ \bibinfo
  {author} {\bibfnamefont {Y.}~\bibnamefont {Wang}},\ }\href {\doibase
  10.1063/1.3570638} {\bibfield  {journal} {\bibinfo  {journal} {Physics of
  Plasmas}\ }\textbf {\bibinfo {volume} {18}},\ \bibinfo {eid} {032705}
  (\bibinfo {year} {2011})}\BibitemShut {NoStop}%
\bibitem [{\citenamefont {Liu}\ \emph {et~al.}(2012)\citenamefont {Liu},
  \citenamefont {Hao}, \citenamefont {Xiang}, \citenamefont {Zhu},
  \citenamefont {Zheng}, \citenamefont {Cao},\ and\ \citenamefont
  {He}}]{Liu_2012PPCF}%
  \BibitemOpen
  \bibfield  {author} {\bibinfo {author} {\bibfnamefont {Z.~J.}\ \bibnamefont
  {Liu}}, \bibinfo {author} {\bibfnamefont {L.}~\bibnamefont {Hao}}, \bibinfo
  {author} {\bibfnamefont {J.}~\bibnamefont {Xiang}}, \bibinfo {author}
  {\bibfnamefont {S.~P.}\ \bibnamefont {Zhu}}, \bibinfo {author} {\bibfnamefont
  {C.~Y.}\ \bibnamefont {Zheng}}, \bibinfo {author} {\bibfnamefont {L.~H.}\
  \bibnamefont {Cao}}, \ and\ \bibinfo {author} {\bibfnamefont {X.~T.}\
  \bibnamefont {He}},\ }\href {http://stacks.iop.org/0741-3335/54/i=9/a=095004}
  {\bibfield  {journal} {\bibinfo  {journal} {Plasma Physics and Controlled
  Fusion}\ }\textbf {\bibinfo {volume} {54}},\ \bibinfo {pages} {095004}
  (\bibinfo {year} {2012})}\BibitemShut {NoStop}%
\bibitem [{\citenamefont {Feng}\ \emph
  {et~al.}(2016{\natexlab{a}})\citenamefont {Feng}, \citenamefont {Zheng},
  \citenamefont {Liu}, \citenamefont {Xiao}, \citenamefont {Wang},\ and\
  \citenamefont {He}}]{Feng_2016POP}%
  \BibitemOpen
  \bibfield  {author} {\bibinfo {author} {\bibfnamefont {Q.~S.}\ \bibnamefont
  {Feng}}, \bibinfo {author} {\bibfnamefont {C.~Y.}\ \bibnamefont {Zheng}},
  \bibinfo {author} {\bibfnamefont {Z.~J.}\ \bibnamefont {Liu}}, \bibinfo
  {author} {\bibfnamefont {C.~Z.}\ \bibnamefont {Xiao}}, \bibinfo {author}
  {\bibfnamefont {Q.}~\bibnamefont {Wang}}, \ and\ \bibinfo {author}
  {\bibfnamefont {X.~T.}\ \bibnamefont {He}},\ }\href {\doibase
  10.1063/1.4960292} {\bibfield  {journal} {\bibinfo  {journal} {Physics of
  Plasmas}\ }\textbf {\bibinfo {volume} {23}},\ \bibinfo {eid} {082106}
  (\bibinfo {year} {2016}{\natexlab{a}})}\BibitemShut {NoStop}%
\bibitem [{\citenamefont {Feng}\ \emph
  {et~al.}(2016{\natexlab{b}})\citenamefont {Feng}, \citenamefont {Xiao},
  \citenamefont {Wang}, \citenamefont {Zheng}, \citenamefont {Liu},
  \citenamefont {Cao},\ and\ \citenamefont {He}}]{Feng_2016PRE}%
  \BibitemOpen
  \bibfield  {author} {\bibinfo {author} {\bibfnamefont {Q.~S.}\ \bibnamefont
  {Feng}}, \bibinfo {author} {\bibfnamefont {C.~Z.}\ \bibnamefont {Xiao}},
  \bibinfo {author} {\bibfnamefont {Q.}~\bibnamefont {Wang}}, \bibinfo {author}
  {\bibfnamefont {C.~Y.}\ \bibnamefont {Zheng}}, \bibinfo {author}
  {\bibfnamefont {Z.~J.}\ \bibnamefont {Liu}}, \bibinfo {author} {\bibfnamefont
  {L.~H.}\ \bibnamefont {Cao}}, \ and\ \bibinfo {author} {\bibfnamefont
  {X.~T.}\ \bibnamefont {He}},\ }\href {\doibase 10.1103/PhysRevE.94.023205}
  {\bibfield  {journal} {\bibinfo  {journal} {Phys. Rev. E}\ }\textbf {\bibinfo
  {volume} {94}},\ \bibinfo {pages} {023205} (\bibinfo {year}
  {2016}{\natexlab{b}})}\BibitemShut {NoStop}%
\bibitem [{\citenamefont {Chapman}\ \emph {et~al.}(2013)\citenamefont
  {Chapman}, \citenamefont {Berger}, \citenamefont {Brunner},\ and\
  \citenamefont {Williams}}]{Chapman_2013PRL}%
  \BibitemOpen
  \bibfield  {author} {\bibinfo {author} {\bibfnamefont {T.}~\bibnamefont
  {Chapman}}, \bibinfo {author} {\bibfnamefont {R.~L.}\ \bibnamefont {Berger}},
  \bibinfo {author} {\bibfnamefont {S.}~\bibnamefont {Brunner}}, \ and\
  \bibinfo {author} {\bibfnamefont {E.~A.}\ \bibnamefont {Williams}},\ }\href
  {\doibase 10.1103/PhysRevLett.110.195004} {\bibfield  {journal} {\bibinfo
  {journal} {Phys. Rev. Lett.}\ }\textbf {\bibinfo {volume} {110}},\ \bibinfo
  {pages} {195004} (\bibinfo {year} {2013})}\BibitemShut {NoStop}%
\bibitem [{\citenamefont {Berger}\ \emph {et~al.}(2013)\citenamefont {Berger},
  \citenamefont {Brunner}, \citenamefont {Chapman}, \citenamefont {Divol},
  \citenamefont {Still},\ and\ \citenamefont {Valeo}}]{Berger_2013POP}%
  \BibitemOpen
  \bibfield  {author} {\bibinfo {author} {\bibfnamefont {R.~L.}\ \bibnamefont
  {Berger}}, \bibinfo {author} {\bibfnamefont {S.}~\bibnamefont {Brunner}},
  \bibinfo {author} {\bibfnamefont {T.}~\bibnamefont {Chapman}}, \bibinfo
  {author} {\bibfnamefont {L.}~\bibnamefont {Divol}}, \bibinfo {author}
  {\bibfnamefont {C.~H.}\ \bibnamefont {Still}}, \ and\ \bibinfo {author}
  {\bibfnamefont {E.~J.}\ \bibnamefont {Valeo}},\ }\href {\doibase
  10.1063/1.4794346} {\bibfield  {journal} {\bibinfo  {journal} {Physics of
  Plasmas}\ }\textbf {\bibinfo {volume} {20}},\ \bibinfo {eid} {032107}
  (\bibinfo {year} {2013})}\BibitemShut {NoStop}%
\end{thebibliography}%

\end{document}